\documentclass[12pt,aps,prd,superscriptaddress,showpacs,longbibliography,floatfix,nofootinbib]{revtex4-1}

\usepackage[utf8]{inputenc}
\usepackage{slashed}
\pdfoutput=1

\usepackage{color}
\usepackage{graphicx}   
\usepackage{bm}
\usepackage{amsmath}
\usepackage{amsfonts}
\usepackage{hyperref}

\def\MS{\overline{\rm MS}}

\begin{document}

\title{Exact $\beta$-function of Yang-Mills theory in 2+1 dimensions}

\author{Paul Romatschke}
\affiliation{Department of Physics, University of Colorado, Boulder, Colorado 80309, USA}
\affiliation{Center for Theory of Quantum Matter, University of Colorado, Boulder, Colorado 80309, USA}

\begin{abstract}
To set the stage, I discuss the $\beta$-function of the massless O(N) model in three dimensions, which can be calculated exactly in the large N limit. Then, I consider SU(N) Yang-Mills theory in 2+1 space-time dimensions. Relating the $\beta$-function to the expectation value of the action in lattice gauge theory, and the latter to the trace of the energy-momentum tensor, I show that $\frac{d \ln g^2/\mu}{d\ln \mu}=-1$ for all $g$ and all N in one particular renormalization scheme. As a consequence, I find that the Yang-Mills $\beta$-function in three dimensions must have the same sign for all finite and positive bare coupling parameters in any renormalization scheme, and all non-trivial infrared fixed points are unreachable in practice.
\end{abstract}

\maketitle


\section{Introduction}

This article once had an introduction. However, I was told that my introduction would get me thrown down the stairs. I don't want to be thrown down the stairs, hence I clipped the introduction.

The calculation will have to stand on its own. I'm sure you will be able to place it into context.

\section{Warm-Up: Critical O(N) Model in D=3, 4 dimensions}
\label{sec:on}

As a warm-up, let me consider the O(N) model with quartic interaction in $D$ dimensions. Finite temperature $T$ is introduced by considering imaginary time  $x_0$ that is compactified on a circle of radius $T^{-1}$. The Euclidean action then reads
\begin{equation}
  \label{eq:action1}
  S_E=\int d^Dx \left[\frac{1}{2}\left(\partial_\mu \vec{\phi}\right)\cdot\left(\partial_\mu \vec{\phi}\right)+m_0^2 \vec{\phi}^2+\frac{\lambda_0}{N}\left(\vec{\phi}^2\right)^2\right]\,,
\end{equation}
where the coupling $\lambda_0$ is dimensionless in $D=4$, but has mass dimension one in $D=3$. The partition function is given by the path integral $Z=\int {\cal D}\phi e^{-S_E}$ which can be conveniently re-written as $\left(\vec{\phi}^2\right)^2=\sigma^2$ using the auxiliary fields $\sigma,\zeta$:
$$
1=\int {\cal D}\sigma \delta\left(\sigma-\vec{\phi}^2\right)=\int {\cal D}{\sigma}{\cal D}\zeta e^{i\int \zeta\left(\sigma-\vec{\phi}^2\right)}\,.
$$
$\sigma$ may be integrated out exactly, leaving only $\zeta$. In the large N limit, only the zero mode $\bar{\zeta}$ contributes to the partition function. As a consequence, the dependence of the action on the scalar fields $\vec{\phi}$ is quadratic, and the path integral over $\vec{\phi}$ may be performed as
\begin{equation}
  \int {\cal D}\phi e^{-\int d^Dx \left[-\frac{1}{2}\vec{\phi} \partial_\mu \partial_\mu \vec{\phi}+i \bar\zeta \vec{\phi}^2+m_0^2 \vec{\phi}^2\right]}=e^{-\frac{N V}{T}\frac{T }{2} \sum_n\int_{{\bf k}} \ln \left(\omega_n^2+{\bf k}^2+m_0^2+2 i \bar \zeta \right)}\,,
  \end{equation}
where $\omega_n=2 \pi n T$, $n\in {\mathbb Z}$ are the Matsubara frequencies and $V$ is the $D-1$ dimensional ``volume'' of the space-time.

Integration over the remaining zero mode $\bar \zeta$ and these may be evaluated exactly from the saddle point of the action. Writing $2 i \bar \zeta=z^*$, the saddle point is given by 
\begin{equation}
  \label{eq:saddle}
  \frac{z^*}{\lambda_0} = 4 T\sum_n \int \frac{d^{D-1}{\bf k}}{(2 \pi)^{D-1}} \frac{1}{\omega_n^2+{\bf k}^2+m_0^2+z^*}=2\int \frac{d^{D-1}{\bf k}}{(2 \pi)^{D-1}} \frac{1+2 n_B(\sqrt{{\bf k}^2+m_0^2+z^*})}{\sqrt{{\bf k}^2+z^*+m_0^2}}\,,
  \end{equation}
where $n_B(x)=\frac{1}{e^{x/T}-1}$ is the thermal distribution for bosons.

Let me first discuss what happens in dimensional regularization, which is the de-facto standard regularization scheme for analytic calculations. In this case, accessing the massless (critical) theory is easy and simply amounts to setting $m_0=0$. In dimensional regularization, one finds
\begin{equation}
  \label{eq:integral}
  \int \frac{d^{D-1}k}{(2 \pi)^{D-1}} \frac{1}{\sqrt{k^2+z^*}}=\frac{1}{(4 \pi)^{\frac{D-1}{2}}} \frac{\Gamma\left(\frac{2-D}{2}\right)}{\Gamma\left(\frac{1}{2}\right)} \left(z^*\right)^{\frac{D-2}{2}}=\left\{\begin{array}{cc}
  -\frac{z^*}{8\pi^2 \epsilon}+{\cal O}(1),& D=4-2\epsilon\\
  -\frac{\sqrt{z^*}}{2\pi}+{\cal O}(\epsilon), & D=3-2\epsilon
\end{array}\right.
  \,,
\end{equation}
which is well-known result in large N field theory \cite{Moshe:2003xn}.

\subsection*{D=4 in $\overline{\rm MS}$: The Wilson-Fisher fixed point}

Therefore, in $D=4-2\epsilon$, the integral (\ref{eq:integral}) has a pole for $\epsilon\rightarrow 0$, and the saddle point condition reads (see e.g. Ref.~\cite{Romatschke:2019gck} for details)
\begin{equation}
\label{eq:saddle1}
0=z^*\left(\frac{1}{\lambda_0}+\frac{1}{4\pi^2 \epsilon}\right)+{\rm finite\ \&T-dependent}\,,\quad D=4-2\epsilon\,.
\end{equation}
 In order for the action (\ref{eq:action1}) to be dimensionless in $D=4-2\epsilon$, $\lambda_0$ must have mass dimension $2 \epsilon$. Thus writing
\begin{equation}
  \label{eq:scaling}
  \frac{1}{\lambda_0}=\frac{Z \mu^{D-4}}{\lambda_R(\mu)}\,,
\end{equation}
where $\mu$ is the renormalization point in the $\MS$ scheme and $\lambda_R$ is the renormalized coupling, the $Z$-factor in $\overline{\rm MS}$ needs to be chosen as $Z=1-\frac{\lambda_R}{4\pi^2 \epsilon}$ in order for the saddled point condition (\ref{eq:saddle1}) to be finite for all temperatures. Since $\lambda_0$ does not depend on the scale $\mu$, it follows that \begin{equation}
  \label{eq:beta4}
  \mu \frac{\partial \lambda_R^{-1}}{\partial \mu}=\frac{2 \epsilon}{\lambda_R}-\frac{1}{2\pi^2}\,,\quad \beta_{\overline{MS}}(\lambda_R)\equiv \frac{\partial \lambda_R}{\partial \ln \mu}=\lambda_R\left(\frac{\lambda_R}{2\pi^2}-2\epsilon\right)\,,\quad D=4-2\epsilon\,\quad {\rm in\ }{\overline{\rm MS}}\,.
\end{equation}
As a consequence, $\beta_{\MS}$ has two zeros (fixed points), namely the free theory point located at $\lambda_R=0$ and the Wilson-Fisher fixed point located at $\lambda_R=4 \pi^2 \epsilon$  \cite{Wilson:1971dc}. This result is exact in the large N limit.  Note that at the Wilson-Fisher fixed point $Z=0$ such that the bare coupling $\lambda_0$ is singular. This implies that for finite and positive-definite bare coupling $\lambda_0<\infty$, the Wilson-Fisher fixed  point can never be reached. In some sense, the Wilson-Fisher fixed point is more of an endpoint, since for $\lambda_0 \in [0,\infty)$ the renormalized coupling only takes on values $\lambda_R\in [0,4 \pi^2 \epsilon)$. In the limit $\epsilon\rightarrow 0$, this shows that the theory is trivial (but the theory is perfectly well defined as a cut-off theory when keeping $\epsilon$ finite).
    
\subsection*{D=3 in $\overline{\rm MS}$: No apparent Wilson-Fisher fixed point}

However, the situation encountered for $D=3-2\epsilon$ dimensions is very different. In this case, the integral (\ref{eq:integral}) does not contain a logarithmic divergence, so the saddle-point condition is finite without renormalization \cite{Romatschke:2019ybu}. Hence $Z=1$ in $\overline{\rm MS}$ and one finds
\begin{equation}
  \label{eq:betatriv}
  \mu \frac{\partial \lambda_R^{-1}}{\partial \mu}=\frac{1}{\lambda_R}\,,\quad
  \beta_{\MS}(\lambda_R)=-\lambda_R\,,\quad D=3-2\epsilon\quad {\rm in\ }{\overline{\rm MS}}\,.
\end{equation}
This result is exact in the large N limit. The running of the coupling is trivially given by the mass dimension of $\lambda$\footnote{As a consequence, the theory is invariant under the combined scale transformations $x^\mu\rightarrow x^\mu/\alpha$, $\lambda_0\rightarrow \alpha\times \lambda_0$ with $\alpha$ the scale parameter. Note that this does \textit{not} imply scale  invariance of the theory, because the bare coupling needs to be rescaled in addition to the space-time coordinates, cf. Refs.~\cite{Komargodski:2011vj,Komargodski:2011xv}.}. The $\beta$ function has a free theory fixed point located at $\lambda_R=0$ in D=3 in the $\overline{\rm MS}$ renormalization scheme. For finite values of $\lambda_R$ (corresponding to finite values of the bare coupling $\lambda_0\in[0,\infty)$), there is no apparent Wilson-Fisher fixed point or any other non-trivial infrared fixed point to leading order in large N. However, in the limit where $\lambda_R\rightarrow \infty$, the derivative $\frac{\partial \lambda_R^{-1}}{\partial \ln \mu}\rightarrow 0$. This ``fixed point'' at $\lambda_R\rightarrow \infty=\lambda^*$ can be used to calculate a critical exponent
  \begin{equation}
    \label{eq:o1}
\omega=\left.\frac{\partial}{\partial \lambda_R^{-1}} \frac{\partial \lambda_R^{-1}}{\partial \ln \mu}\right|_{\lambda^*}=1\,.
  \end{equation}
(Note that since $\beta_{\MS}$ does not possess a fixed-point for finite coupling, it can not be used to define $\omega$ in the $\MS$ scheme).

It has been brought to my attention that some readers may not trust results derived using dimensional regularization. So let me investigate what happens to (\ref{eq:saddle}) when using cut-off regularization instead of dimensional regularization for $D=3$. For the case of cut-off regularization, I will \textit{not} set the bare mass parameter $m_0$ to zero, for reasons that will become clear in a moment. The saddle point condition (\ref{eq:saddle}) in cut-off regularization becomes
\begin{equation}
  \label{eq:integral2}
 \frac{z^*}{\lambda_0}=2 \int_0^\Lambda \frac{d k k}{2 \pi} \frac{1}{\sqrt{k^2+z^*+m_0^2}}=\frac{\Lambda}{\pi}-\frac{\sqrt{z^*+m_0^2}}{\pi}+{\cal O}(\Lambda^{-1})+{\rm finite}\,,
\end{equation}
where $\Lambda$ is the momentum cut-off.
Obviously, there is a linear divergence in cut-off regularization. However, the divergence cannot be absorbed into a renormalized coupling, because the only dependence on the coupling constant is in the term $\frac{z^*}{\lambda_0}$. Instead, note that the relevant combination appearing in the propagator is the combination $z^*+m_0^2$, which reads
\begin{equation}
  \label{eq:resolution}
 \frac{z^*+m_0^2}{\lambda_0}=\frac{m^2_R}{\lambda_0}-\frac{\sqrt{z^*+m_0^2}}{\pi}+{\rm finite}\,,
\end{equation}
where I have introduced the renormalized mass $m_R^2(\mu)$ as
\begin{equation}
  m_R^2(\mu)=m_0^2+\frac{\Lambda \lambda_0}{\pi}\,.
\end{equation}
One may tune $m_0^2$ to the critical theory defined by $m_R^2=0$, such that the cut-off regularized saddle point condition becomes identical to the one in dimensional regularization. Therefore, one has again $Z=1$ for the coupling, but a non-trivial running for the renormalized mass. As in dimensional regularization, there are no non-trivial infrared fixed points in the $\beta$-function  in $\overline{\rm MS}$ for $\lambda_0\in[0,\infty)$ at large N.

\subsection*{D=3 in other renormalization schemes: recovering the fixed point?}

There is a school of thought that would have the large N Wilson-Fisher fixed point persists to three dimension by simply using the result (\ref{eq:beta4}) derived for $D=4-2\epsilon$ and sending $\epsilon\rightarrow \frac{1}{2}$ at the end. I find this view somewhat confusing, even though it is to some extent only a semantic difference. Putting $2\epsilon=1$ in a calculation where terms of order ${\cal O}(\epsilon)$ have been neglected is at best an approximation, but not an exact result. This approximation may be useful in many contexts where no other non-perturbative solution exists, but for theories that are exactly solvable in D=3, the exact result (\ref{eq:integral})  does not possess a logarithmic divergence; therefore, using $2\epsilon=1$ in (\ref{eq:beta4}) no longer corresponds to the $\overline{\rm MS}$ renormalization scheme, because non-divergent terms are subtracted.

However, the idea of modifying the renormalization scheme is certainly a valid one: after all, the $\MS$-scheme is just a convention, and one may just as well define a ``renormalization'' scheme by absorbing finite terms in the definition of the coupling. For instance, putting $Z=1-\lambda/c$ in Eq.~(\ref{eq:betatriv}) where $c$ is a finite constant leads to 
\begin{equation}
  \label{eq:betatriv2}
  \mu \frac{\partial \lambda^{-1}}{\partial \mu}=\frac{1}{\lambda}-\frac{1}{c}\,,\quad \beta=\lambda\left(\frac{\lambda}{c}-1\right)\,,\quad D=3-2\epsilon\quad {\rm NOT\ in\ }{\overline{\rm MS}}\,.
\end{equation}

Now there is an apparent non-trivial fixed point in the $\beta$ function located at $\lambda=c$. The critical exponent $\omega=\frac{\partial}{\partial \lambda^{-1}} \frac{\partial \lambda^{-1}}{\partial \ln \mu}=1$ matches the result (\ref{eq:o1}). However, the scheme is no longer $\MS$, so one could object that the fixed point thus found is not associated with a physical property of the system. For this reason, let me discuss a ``physical'' definition of the coupling instead of (\ref{eq:betatriv2}).

\subsection*{``Physical Coupling Definition'' in D=3 and ``unreachable'' fixed points}

\label{subsi}

A popular scheme to define the running coupling of the theory is via a physical process such as two-by-two scattering. In this case, one has to calculate the four-point function of the theory, which can be done exactly in the large N limit. Specifically, one finds for the connected amputated four-point function $\Gamma=\langle \vec{\phi}^2(P) \vec{\phi}^2(-P)\rangle$ \cite{Maldacena:2019}
\begin{eqnarray}
  \Gamma&=&\left[A(P)-\frac{2 \lambda_0}{N} A^2(P)+\left(\frac{2 \lambda_0}{N} A^2(P)\right)^2+\ldots\right]=\frac{A(P)}{1+\frac{2 \lambda_0}{N}A(P)}\,,\nonumber\\
  A(P)&=&2 N T\sum_n \int \frac{d^{2}{\bf k}}{(2 \pi)^{2}}G(K)G(P-K)\,,
\end{eqnarray}
where $G(K)=\frac{1}{\omega_n^2+{\bf k}^2+z^*}$ and $z^*$ is the solution to the gap equation (\ref{eq:saddle}) for $D=3$. In the zero temperature limit,
\begin{equation}
  \lim_{T\rightarrow 0}A(P)=2N \int\frac{d^3K}{(2\pi)^3}G(K)G(P-K)=\frac{N}{4 \sqrt{P^2}}\,.
\end{equation}
Therefore, it is possible to define an effective coupling $\lambda_{\rm eff}(P)$ as
\begin{equation}
  \label{eq:leffdef}
\frac{\lambda_{\rm eff}(P)}{2}=1-\frac{\Gamma}{A(P)}=\frac{\lambda_0}{2 \sqrt{P^2}+\lambda_0}=\frac{\lambda_0}{2 \sqrt{P^2}}-\left(\frac{\lambda_0}{2 \sqrt{P^2}}\right)^2+\ldots
\end{equation}
Putting $\mu=\sqrt{P^2}$, this effective coupling has an effective $\beta$-function, which can be calculated as
\begin{equation}
  \label{eq:fixi}
\mu \frac{\partial \lambda_{\rm eff}(\mu)}{\partial \mu}=-\lambda_{\rm eff}\left(1-\frac{\lambda_{\rm eff}}{2}\right)\,.
\end{equation}
In the large N limit, this result is exact. Taken at face value, the result implies the existence of a non-trivial fixed point located at $\lambda_{\rm eff}=2$. As in the example of Eq.~(\ref{eq:betatriv2}), the renormalization scheme used to define $\lambda_{\rm eff}$ is not $\MS$. It seems that depending on the choice of renormalization scheme, there is or isn't a fixed point in the $\beta$-function. What is going on?

The renormalization group invariant $\beta \frac{\partial}{\partial \lambda}$ implies that under changes of the definition of the coupling $\lambda\rightarrow \lambda^\prime$, the associated $\beta$ function should change as 
\begin{equation}
  \beta(\lambda)=\beta^\prime(\lambda^\prime)  \frac{d \lambda}{d \lambda^\prime}\,.
  \end{equation}
Fixed points are zeros of the $\beta$ function, and the above equation seems to imply that the existence of fixed points (but not their location) is independent from the definition of the coupling. This is generally assumed to be true. However, this assumption is violated if the transformation $\lambda\rightarrow \lambda^\prime$ is not invertible, such that the Jacobian $\frac{d \lambda}{d \lambda^\prime}$ becomes singular. The condition for the Jacobian to be regular for all couplings is that the transformation of the coupling is analytic. Is is easy to verify for the case at hand that this is \textit{not} the case: taking $\lambda_R=\frac{\lambda_0}{\mu}$ from (\ref{eq:scaling}) and $\lambda_{\rm eff}$ from (\ref{eq:leffdef}), one has
\begin{equation}
  \lambda_R=\frac{\lambda_{\rm eff}}{1-\frac{\lambda_{\rm eff}}{2}}\,,
\end{equation}
which is singular at $\lambda_{\rm eff}=2$. Note that the location of the singularity in the transformation of the coupling precisely matches the apparent fixed-point found for the effective $\beta$-function in (\ref{eq:fixi}). Because $\lambda_R$ diverges for $\lambda_{\rm eff}\rightarrow 2$, the apparent fixed point at $\lambda_{\rm eff}=2$ cannot be reached in practice by any positive and finite bare coupling $\lambda_0/\mu \in [0,\infty]$. It is therefore unreachable in practice, and hence the $\beta$-function in the ``physical coupling definition'' has the same sign as $\beta_{\MS}$ in (\ref{eq:betatriv}) for all positive and finite bare coupling values.

  \subsection*{Lessons from the O(N) model in D=3}

  In all cases discussed above for the massless O(N) model at large N, there is no zero of the $\beta$-function for bare couplings $\lambda_0\in[0,\infty)$, regardless of the renormalization scheme, regularization scheme, or coupling definition. Depending on the scheme, apparent zeros of the $\beta$-function may be found, but these fixed points can not be reached with any finite bare coupling parameter. While infrared fixed points in the O(N) model may arise once subleading corrections in 1/N are taken into account, the above findings are exact in the large N limit and provide the interpretive stage for the following calculations in Yang-Mills theory.


\section{Yang-Mills Theory in 2+1 dimensions}

The Lagrangian density for Yang-Mills theory in 2+1 dimensional Minkowski space-time in continuum formulation is given by
\begin{equation}
  \label{eq:ll}
  {\cal L}=-\frac{1}{4} F_{\mu\nu}^a F^{\mu\nu a}\,,\quad F_{\mu\nu}=\partial_\mu A_\nu^a-\partial_\nu A^a_\mu+ g f^{abc}A_\mu^b A_\nu^c\,,
\end{equation}
where $A^a_\mu$ with $a=1,2,\ldots N^2-1$ is the SU(N) gauge field, $f^{abc}$ are the SU(N) structure constants and $g$ is the (mass dimension one) Yang-Mills coupling. The theory is put at finite temperature using the standard construction in the imaginary time formalism, e.g. by performing the replacement \cite{Laine:2016hma}
\begin{equation}
  \label{eq:rot}
  t\rightarrow -i x_0\,,
\end{equation}
such that
\begin{equation}
i S=i \int dt d^2x {\cal L}\rightarrow - \int d^3x \frac{1}{4} F_{\mu\nu}^aF_{\mu\nu}^a=-S_E\,,
\end{equation}
where $\mu,\nu$ are now indices in three dimensional Euclidean space and $S_E$ is the Euclidean action. In the continuum formulation, gauge invariance implies the presence of flat directions when integrating over the gauge fields $A^\mu$, which would render a naive expression for the partition function divergent. A non-perturbative formulation of the partition function is found by replacing the continuum fields $A_\mu(x)$ by link variables $U_\mu(x)=e^{-i g a A^a_\mu(x) T^a}$ where $T^a$ are the generators of SU(N) and space-time has been discretized as a cube with lattice spacing $a$. The discretized Euclidean action then becomes \cite{Montvay:1994cy}
\begin{equation}
  S_E=\frac{2 N}{g^2 a} \sum_{x}\sum_{\Box}\left(1-\frac{1}{N}{\rm Re}{\rm Tr} U_\Box\right)\,,\quad U_\Box=U_i(x)U_j(x+i)U_i^\dagger(x+j)U^\dagger_j(x)\,,
\end{equation}
where the sum is over all sites $x$ of the three-dimensional lattice and over all orientations of plaquettes $U_\Box$. The lattice has periodic boundary condition in all directions, and the length of the time-like direction of the lattice is related to the temperature as $T=\frac{1}{N_\tau a}$, where $N_\tau$ are the number of lattice sites in the temporal direction. In the lattice formulation, the partition function is given by
\begin{equation}
  Z=\int \prod_x dU(x) e^{-S_E}\,,
\end{equation}
which lends itself to numerical Monte-Carlo simulations via importance sampling.

For the case at hand, it is sufficient to recall that the partition function is related to the pressure $p$ of Yang-Mills theory as
\begin{equation}
  p=\frac{T}{V}\ln Z\,,
\end{equation}
where $V$ is again the ``volume'' of two-dimensional space.
A standard operation in lattice gauge theory thermodynamics is to calculate the derivative of $p$ with respect to the lattice coupling $\beta_L\equiv \frac{2 N}{g^2 a}$, e.g.
\begin{equation}
  \label{eq:id2}
  \frac{\partial p/T^3}{\partial \beta_L}=\frac{1}{V T^2} \frac{\int \prod_x dU(x) e^{-S_E} (-\frac{\partial S_E}{\partial \beta_L})}{Z}=-\frac{1}{V T^2}\frac{\langle S_E \rangle}{\beta_L} \,,
\end{equation}
where $\langle S_E\rangle$ is the thermodynamic expectation value of the action, and I have divided the pressure by $T^3$ to make both sides dimensionless. Using
$$
\frac{\partial p/T^3}{\partial \beta_L}=\frac{\partial p/T^3}{\partial T} \left(\frac{\partial \beta_L}{\partial T}\right)^{-1}\,,
  $$
writing $g^2 a=g^2/(T N_\tau)$, and keeping $N_\tau$ fixed (but large) leads to
\begin{equation}
  \frac{\partial p/T^3}{\partial T}=\frac{1}{V T^2}\langle S_E \rangle \frac{\partial \ln (g^2/T)}{\partial T}\,.
\end{equation}

Basic thermodynamic relations govern the relation between the pressure $p$, the entropy density $s$ and the energy density $\epsilon$ such as
\begin{equation}
  \frac{\epsilon+p}{T}=s=\frac{\partial p}{\partial T}\,.
  \end{equation}
Using these one finds 
\begin{equation}
  \label{bla}
  \frac{\partial p/T^3}{\partial T}=\frac{\epsilon-2 p}{T^4}\,.
\end{equation}
However, in thermodynamic equilibrium, the expectation value of the energy-momentum tensor is related to $\epsilon,p$ as $\langle T^{\mu\nu}\rangle={\rm diag}\left(\epsilon,p,p\right)$ such that $\epsilon-2 p =\langle T^\mu_\mu\rangle$. Therefore, I find the relation
\begin{equation}
  \label{reli}
  \langle T^\mu_\mu\rangle = \frac{T^2}{V}\langle S_E \rangle \frac{\partial \ln (g^2/T)}{\partial T}\,,
\end{equation}
which is exact and corresponds to the well-known four-dimensional result \cite{Montvay:1994cy} in $D=3$. The above relation may be used to calculate the $\beta$-function numerically by means of lattice gauge theory simulation, for instance by calculating the expectation value of the action $\langle S_E\rangle$ and similarly calculating the trace of the energy-momentum tensor $\langle T^\mu_\mu\rangle$ e.g. via gradient flow \cite{Kitazawa:2017qab}. However, for the case of three dimensions, such calculations may be unnecessary, as I will argue below.

It is straightforward to derive the energy-momentum tensor from the Minkowski Lagrangian (\ref{eq:ll}), finding \cite{Peskin:1995ev}
\begin{equation}
  T^{\mu\nu}=-F^{\mu \lambda a} F^{\nu a}_{\ \lambda}+\frac{1}{4}g^{\mu\nu} F_{\rho \lambda}^a F^{\rho \lambda a}\,.
\end{equation}
The trace of the energy momentum tensor in $D=3$ dimensions is given by
\begin{equation}
  T^\mu_\mu=-\frac{1}{4} F_{\rho \lambda}^a F^{\rho \lambda a}={\cal L}\,.
\end{equation}
This relation differs from its equivalent in four dimensions. In three dimensions, the trace of the energy-momentum tensor does not vanish. 
Moreover, rotating into Euclidean time (\ref{eq:rot}), the trace of the energy-momentum tensor is related to the Euclidean action as 
\begin{equation}
  \label{eq:resl}
\int d^3x T^\mu_\mu=-S_E=-\int d^3x \frac{1}{4} F_{\mu\nu}^a F_{\mu\nu}^a\,.
\end{equation}
As a consequence, the expectation value of the action in the exact relation (\ref{reli}) may be replaced by the corresponding expectation value of $T^\mu_\mu$. While classical, this relation implies that the same operator -- discretized on the lattice -- controls the action and the trace of the energy momentum tensor. Since the same operator can be used for both, a lattice simulation will give the same expectation value for both quantities, hence the relation (\ref{eq:resl}) can be expected to hold outside the classical regime. It is of crucial importance to note that the replacement (\ref{eq:resl}) does not work in $D=4$ because there the classical Lagrangian is not proportional to the trace of the energy-momentum tensor.

Eq.~(\ref{eq:resl}) does not  depend on any details of a lattice discretization. Hence both the continuum and infinite volume limit may be taken, such that $\langle S_E \rangle$ may be evaluated in the continuum theory where
\begin{equation}
  \langle S_E\rangle=-\int d^3x \langle T^\mu_\mu\rangle = - \frac{V}{T} \langle T^\mu_\mu\rangle\,.
  \end{equation}
 From this equality, I find
\begin{equation}
\frac{\partial \ln (g^2/T)}{\partial \ln T}=-1\,,\quad {\rm or}\quad \frac{\partial \ln (g^2 a)}{\partial \ln a}=1\,.
  \label{eq:norun}
\end{equation}
This relation is exact for all $g$ and all N, but depends on the choice of scheme.

\section{Discussion}

Eq.~(\ref{eq:norun}) implies that the dimensionless coupling $g^2/T$ runs trivially, e.g. there are no non-trivial infrared fixed points for any positive and finite bare lattice coupling $\beta_L$ in three dimensional Yang-Mills theory.  The situation is similar to the O(N) model in $D=3$ encountered in section \ref{sec:on}, but unlike the O(N) model, Eq.~(\ref{eq:norun}) is exact for all N.

It is possible to use a different definition $g^\prime$ for the coupling, e.g. via the scattering amplitude, which will possess a non-trivial fixed point at $g^\prime=g^*$. However, similar to the example of the O(N) model discussed in section \ref{sec:on}, in view of (\ref{eq:norun}) this fixed point must correspond to a singular Jacobian for the transformation $g\rightarrow g^\prime$. Therefore, it is an unreachable fixed point in the sense that $g^\prime<g^*$ for any positive and finite lattice coupling $\beta_L\in [0,\infty)$.

  While employing the scattering amplitude in order to define the lattice spacing $a$ through the lattice coupling $\beta_L$ is a perfectly well-defined prescription, in three dimensional Yang-Mills theory there is an alternative option: just use the bare coupling. As in the case of a true conformal field theory, one may define a fiducial scale (say $g_0^2=1$ GeV) after which the physical lattice spacing is given by
  \begin{equation}
\label{eq:quick}
    a=\frac{2 N}{\beta_L g_0^2}\,.
  \end{equation}
  In light of (\ref{eq:norun}) this relation is exact and has been employed in Monte Carlo simulations of lower-dimensional gauge theories motivated by string theory, cf. Refs.~\cite{Aharony:2004ig,Catterall:2010fx,Hanada:2016qbz}. Using (\ref{eq:quick}), scale setting is trivial and does not require the calculation of the string tension or flowed quantities. Of course, both procedures are correct in three dimensional Yang-Mills theory, but only the one from Eq.~(\ref{eq:quick}) is effortless.

  \section{Acknowledgments}

  This work was supported by the Department of Energy, DOE award No DE-SC0017905.  I would like to thank A.~Carroso, S.P.~deAlwis, T.~DeGrand, A.~Hasenfratz, W.~Jay, Z.~Komargodski, E.~Neil, L.~Radzihovsky, B.~Svetitsky and L.~Yaffe for discussions and helpful comments on this manuscript.

\bibliography{beta}

\begin{thebibliography}{14}%
\makeatletter
\providecommand \@ifxundefined [1]{%
 \@ifx{#1\undefined}
}%
\providecommand \@ifnum [1]{%
 \ifnum #1\expandafter \@firstoftwo
 \else \expandafter \@secondoftwo
 \fi
}%
\providecommand \@ifx [1]{%
 \ifx #1\expandafter \@firstoftwo
 \else \expandafter \@secondoftwo
 \fi
}%
\providecommand \natexlab [1]{#1}%
\providecommand \enquote  [1]{``#1''}%
\providecommand \bibnamefont  [1]{#1}%
\providecommand \bibfnamefont [1]{#1}%
\providecommand \citenamefont [1]{#1}%
\providecommand \href@noop [0]{\@secondoftwo}%
\providecommand \href [0]{\begingroup \@sanitize@url \@href}%
\providecommand \@href[1]{\@@startlink{#1}\@@href}%
\providecommand \@@href[1]{\endgroup#1\@@endlink}%
\providecommand \@sanitize@url [0]{\catcode `\\12\catcode `\$12\catcode
  `\&12\catcode `\#12\catcode `\^12\catcode `\_12\catcode `\%12\relax}%
\providecommand \@@startlink[1]{}%
\providecommand \@@endlink[0]{}%
\providecommand \url  [0]{\begingroup\@sanitize@url \@url }%
\providecommand \@url [1]{\endgroup\@href {#1}{\urlprefix }}%
\providecommand \urlprefix  [0]{URL }%
\providecommand \Eprint [0]{\href }%
\providecommand \doibase [0]{http://dx.doi.org/}%
\providecommand \selectlanguage [0]{\@gobble}%
\providecommand \bibinfo  [0]{\@secondoftwo}%
\providecommand \bibfield  [0]{\@secondoftwo}%
\providecommand \translation [1]{[#1]}%
\providecommand \BibitemOpen [0]{}%
\providecommand \bibitemStop [0]{}%
\providecommand \bibitemNoStop [0]{.\EOS\space}%
\providecommand \EOS [0]{\spacefactor3000\relax}%
\providecommand \BibitemShut  [1]{\csname bibitem#1\endcsname}%
\let\auto@bib@innerbib\@empty
\bibitem [{\citenamefont {Moshe}\ and\ \citenamefont
  {Zinn-Justin}(2003)}]{Moshe:2003xn}%
  \BibitemOpen
  \bibfield  {author} {\bibinfo {author} {\bibfnamefont {Moshe}\ \bibnamefont
  {Moshe}}\ and\ \bibinfo {author} {\bibfnamefont {Jean}\ \bibnamefont
  {Zinn-Justin}},\ }\bibfield  {title} {\enquote {\bibinfo {title} {{Quantum
  field theory in the large N limit: A Review}},}\ }\href {\doibase
  10.1016/S0370-1573(03)00263-1} {\bibfield  {journal} {\bibinfo  {journal}
  {Phys. Rept.}\ }\textbf {\bibinfo {volume} {385}},\ \bibinfo {pages}
  {69--228} (\bibinfo {year} {2003})},\ \Eprint
  {http://arxiv.org/abs/hep-th/0306133} {arXiv:hep-th/0306133 [hep-th]}
  \BibitemShut {NoStop}%
\bibitem [{\citenamefont
  {Romatschke}(2019{\natexlab{a}})}]{Romatschke:2019gck}%
  \BibitemOpen
  \bibfield  {author} {\bibinfo {author} {\bibfnamefont {Paul}\ \bibnamefont
  {Romatschke}},\ }\bibfield  {title} {\enquote {\bibinfo {title} {{Analytic
  Transport from Weak to Strong Coupling in the O(N) model}},}\ }\href@noop {}
  {\  (\bibinfo {year} {2019}{\natexlab{a}})},\ \Eprint
  {http://arxiv.org/abs/1905.09290} {arXiv:1905.09290 [hep-th]} \BibitemShut
  {NoStop}%
\bibitem [{\citenamefont {Wilson}\ and\ \citenamefont
  {Fisher}(1972)}]{Wilson:1971dc}%
  \BibitemOpen
  \bibfield  {author} {\bibinfo {author} {\bibfnamefont {Kenneth~G.}\
  \bibnamefont {Wilson}}\ and\ \bibinfo {author} {\bibfnamefont {Michael~E.}\
  \bibnamefont {Fisher}},\ }\bibfield  {title} {\enquote {\bibinfo {title}
  {{Critical exponents in 3.99 dimensions}},}\ }\href {\doibase
  10.1103/PhysRevLett.28.240} {\bibfield  {journal} {\bibinfo  {journal} {Phys.
  Rev. Lett.}\ }\textbf {\bibinfo {volume} {28}},\ \bibinfo {pages} {240--243}
  (\bibinfo {year} {1972})}\BibitemShut {NoStop}%
\bibitem [{\citenamefont
  {Romatschke}(2019{\natexlab{b}})}]{Romatschke:2019ybu}%
  \BibitemOpen
  \bibfield  {author} {\bibinfo {author} {\bibfnamefont {Paul}\ \bibnamefont
  {Romatschke}},\ }\bibfield  {title} {\enquote {\bibinfo {title}
  {{Finite-Temperature Conformal Field Theory Results for All Couplings: O(N)
  Model in 2+1 Dimensions}},}\ }\href {\doibase 10.1103/PhysRevLett.122.231603}
  {\bibfield  {journal} {\bibinfo  {journal} {Phys. Rev. Lett.}\ }\textbf
  {\bibinfo {volume} {122}},\ \bibinfo {pages} {231603} (\bibinfo {year}
  {2019}{\natexlab{b}})},\ \Eprint {http://arxiv.org/abs/1904.09995}
  {arXiv:1904.09995 [hep-th]} \BibitemShut {NoStop}%
\bibitem [{\citenamefont {Komargodski}\ and\ \citenamefont
  {Schwimmer}(2011)}]{Komargodski:2011vj}%
  \BibitemOpen
  \bibfield  {author} {\bibinfo {author} {\bibfnamefont {Zohar}\ \bibnamefont
  {Komargodski}}\ and\ \bibinfo {author} {\bibfnamefont {Adam}\ \bibnamefont
  {Schwimmer}},\ }\bibfield  {title} {\enquote {\bibinfo {title} {{On
  Renormalization Group Flows in Four Dimensions}},}\ }\href {\doibase
  10.1007/JHEP12(2011)099} {\bibfield  {journal} {\bibinfo  {journal} {JHEP}\
  }\textbf {\bibinfo {volume} {12}},\ \bibinfo {pages} {099} (\bibinfo {year}
  {2011})},\ \Eprint {http://arxiv.org/abs/1107.3987} {arXiv:1107.3987
  [hep-th]} \BibitemShut {NoStop}%
\bibitem [{\citenamefont {Komargodski}(2012)}]{Komargodski:2011xv}%
  \BibitemOpen
  \bibfield  {author} {\bibinfo {author} {\bibfnamefont {Zohar}\ \bibnamefont
  {Komargodski}},\ }\bibfield  {title} {\enquote {\bibinfo {title} {{The
  Constraints of Conformal Symmetry on RG Flows}},}\ }\href {\doibase
  10.1007/JHEP07(2012)069} {\bibfield  {journal} {\bibinfo  {journal} {JHEP}\
  }\textbf {\bibinfo {volume} {07}},\ \bibinfo {pages} {069} (\bibinfo {year}
  {2012})},\ \Eprint {http://arxiv.org/abs/1112.4538} {arXiv:1112.4538
  [hep-th]} \BibitemShut {NoStop}%
\bibitem [{\citenamefont {Maldacena}(2019)}]{Maldacena:2019}%
  \BibitemOpen
  \bibfield  {author} {\bibinfo {author} {\bibfnamefont {Juan~Martin}\
  \bibnamefont {Maldacena}},\ }\bibfield  {title} {\enquote {\bibinfo {title}
  {{TASI 2019 lectures on Large N}},}\ }in\ \href@noop {} {\emph {\bibinfo
  {booktitle} {{TASI 2019, Boulder, USA, June 3-28, 2019}}}}\ (\bibinfo {year}
  {2019})\BibitemShut {NoStop}%
\bibitem [{\citenamefont {Laine}\ and\ \citenamefont
  {Vuorinen}(2016)}]{Laine:2016hma}%
  \BibitemOpen
  \bibfield  {author} {\bibinfo {author} {\bibfnamefont {Mikko}\ \bibnamefont
  {Laine}}\ and\ \bibinfo {author} {\bibfnamefont {Aleksi}\ \bibnamefont
  {Vuorinen}},\ }\bibfield  {title} {\enquote {\bibinfo {title} {{Basics of
  Thermal Field Theory}},}\ }\href {\doibase 10.1007/978-3-319-31933-9}
  {\bibfield  {journal} {\bibinfo  {journal} {Lect. Notes Phys.}\ }\textbf
  {\bibinfo {volume} {925}},\ \bibinfo {pages} {pp.1--281} (\bibinfo {year}
  {2016})},\ \Eprint {http://arxiv.org/abs/1701.01554} {arXiv:1701.01554
  [hep-ph]} \BibitemShut {NoStop}%
\bibitem [{\citenamefont {Montvay}\ and\ \citenamefont
  {Munster}(1997)}]{Montvay:1994cy}%
  \BibitemOpen
  \bibfield  {author} {\bibinfo {author} {\bibfnamefont {I.}~\bibnamefont
  {Montvay}}\ and\ \bibinfo {author} {\bibfnamefont {G.}~\bibnamefont
  {Munster}},\ }\href {\doibase 10.1017/CBO9780511470783} {\emph {\bibinfo
  {title} {{Quantum fields on a lattice}}}},\ Cambridge Monographs on
  Mathematical Physics\ (\bibinfo  {publisher} {Cambridge University Press},\
  \bibinfo {year} {1997})\BibitemShut {NoStop}%
\bibitem [{\citenamefont {Kitazawa}\ \emph {et~al.}(2017)\citenamefont
  {Kitazawa}, \citenamefont {Iritani}, \citenamefont {Asakawa},\ and\
  \citenamefont {Hatsuda}}]{Kitazawa:2017qab}%
  \BibitemOpen
  \bibfield  {author} {\bibinfo {author} {\bibfnamefont {Masakiyo}\
  \bibnamefont {Kitazawa}}, \bibinfo {author} {\bibfnamefont {Takumi}\
  \bibnamefont {Iritani}}, \bibinfo {author} {\bibfnamefont {Masayuki}\
  \bibnamefont {Asakawa}}, \ and\ \bibinfo {author} {\bibfnamefont {Tetsuo}\
  \bibnamefont {Hatsuda}},\ }\bibfield  {title} {\enquote {\bibinfo {title}
  {{Correlations of the energy-momentum tensor via gradient flow in SU(3)
  Yang-Mills theory at finite temperature}},}\ }\href {\doibase
  10.1103/PhysRevD.96.111502} {\bibfield  {journal} {\bibinfo  {journal} {Phys.
  Rev.}\ }\textbf {\bibinfo {volume} {D96}},\ \bibinfo {pages} {111502}
  (\bibinfo {year} {2017})},\ \Eprint {http://arxiv.org/abs/1708.01415}
  {arXiv:1708.01415 [hep-lat]} \BibitemShut {NoStop}%
\bibitem [{\citenamefont {Peskin}\ and\ \citenamefont
  {Schroeder}(1995)}]{Peskin:1995ev}%
  \BibitemOpen
  \bibfield  {author} {\bibinfo {author} {\bibfnamefont {Michael~E.}\
  \bibnamefont {Peskin}}\ and\ \bibinfo {author} {\bibfnamefont {Daniel~V.}\
  \bibnamefont {Schroeder}},\ }\href
  {http://www.slac.stanford.edu/~mpeskin/QFT.html} {\emph {\bibinfo {title}
  {{An Introduction to quantum field theory}}}}\ (\bibinfo  {publisher}
  {Addison-Wesley},\ \bibinfo {address} {Reading, USA},\ \bibinfo {year}
  {1995})\BibitemShut {NoStop}%
\bibitem [{\citenamefont {Aharony}\ \emph {et~al.}(2004)\citenamefont
  {Aharony}, \citenamefont {Marsano}, \citenamefont {Minwalla},\ and\
  \citenamefont {Wiseman}}]{Aharony:2004ig}%
  \BibitemOpen
  \bibfield  {author} {\bibinfo {author} {\bibfnamefont {Ofer}\ \bibnamefont
  {Aharony}}, \bibinfo {author} {\bibfnamefont {Joe}\ \bibnamefont {Marsano}},
  \bibinfo {author} {\bibfnamefont {Shiraz}\ \bibnamefont {Minwalla}}, \ and\
  \bibinfo {author} {\bibfnamefont {Toby}\ \bibnamefont {Wiseman}},\ }\bibfield
   {title} {\enquote {\bibinfo {title} {{Black hole-black string phase
  transitions in thermal 1+1 dimensional supersymmetric Yang-Mills theory on a
  circle}},}\ }\href {\doibase 10.1088/0264-9381/21/22/010} {\bibfield
  {journal} {\bibinfo  {journal} {Class. Quant. Grav.}\ }\textbf {\bibinfo
  {volume} {21}},\ \bibinfo {pages} {5169--5192} (\bibinfo {year} {2004})},\
  \Eprint {http://arxiv.org/abs/hep-th/0406210} {arXiv:hep-th/0406210 [hep-th]}
  \BibitemShut {NoStop}%
\bibitem [{\citenamefont {Catterall}\ \emph {et~al.}(2010)\citenamefont
  {Catterall}, \citenamefont {Joseph},\ and\ \citenamefont
  {Wiseman}}]{Catterall:2010fx}%
  \BibitemOpen
  \bibfield  {author} {\bibinfo {author} {\bibfnamefont {Simon}\ \bibnamefont
  {Catterall}}, \bibinfo {author} {\bibfnamefont {Anosh}\ \bibnamefont
  {Joseph}}, \ and\ \bibinfo {author} {\bibfnamefont {Toby}\ \bibnamefont
  {Wiseman}},\ }\bibfield  {title} {\enquote {\bibinfo {title} {{Thermal phases
  of D1-branes on a circle from lattice super Yang-Mills}},}\ }\href {\doibase
  10.1007/JHEP12(2010)022} {\bibfield  {journal} {\bibinfo  {journal} {JHEP}\
  }\textbf {\bibinfo {volume} {12}},\ \bibinfo {pages} {022} (\bibinfo {year}
  {2010})},\ \Eprint {http://arxiv.org/abs/1008.4964} {arXiv:1008.4964
  [hep-th]} \BibitemShut {NoStop}%
\bibitem [{\citenamefont {Hanada}\ and\ \citenamefont
  {Romatschke}(2017)}]{Hanada:2016qbz}%
  \BibitemOpen
  \bibfield  {author} {\bibinfo {author} {\bibfnamefont {Masanori}\
  \bibnamefont {Hanada}}\ and\ \bibinfo {author} {\bibfnamefont {Paul}\
  \bibnamefont {Romatschke}},\ }\bibfield  {title} {\enquote {\bibinfo {title}
  {{Lattice Simulations of 10d Yang-Mills toroidally compactified to 1d, 2d and
  4d}},}\ }\href {\doibase 10.1103/PhysRevD.96.094502} {\bibfield  {journal}
  {\bibinfo  {journal} {Phys. Rev.}\ }\textbf {\bibinfo {volume} {D96}},\
  \bibinfo {pages} {094502} (\bibinfo {year} {2017})},\ \Eprint
  {http://arxiv.org/abs/1612.06395} {arXiv:1612.06395 [hep-th]} \BibitemShut
  {NoStop}%
\end{thebibliography}%
\end{document}